\documentclass[prb,amsmath,superscriptaddress,citeautoscript,twocolumn,
showkeys,showpacs,floatfix]{revtex4}
\usepackage{bm}
\usepackage{amssymb}
\usepackage{graphicx}

\begin{document}

\title{Transport through molecular junctions with a nonequilibrium phonon population}

\author{O. Entin-Wohlman}

\altaffiliation{Also at Tel Aviv University, Tel Aviv 69978,
Israel}


\affiliation{Department of Physics and the Ilse Katz Center for
Meso- and Nano-Scale Science and Technology, Ben Gurion
University, Beer Sheva 84105, Israel}

\affiliation{Albert Einstein Minerva Center for Theoretical
Physics, Weizmann Institute of Science, Rehovot 76100, Israel}

\author{Y. Imry}

\affiliation{Department of Condensed Matter Physics,  Weizmann
Institute of Science, Rehovot 76100, Israel}

\author{A. Aharony}

\altaffiliation{Also at Tel Aviv University,
Tel Aviv 69978, Israel}

\affiliation{Department of Physics and the Ilse Katz Center for
Meso- and Nano-Scale Science and Technology, Ben Gurion
University, Beer Sheva 84105, Israel}

\date{\today}

\begin{abstract}

The calculation of the nonlinear conductance of a single-molecule junction is revisited.  The self energy on the junction resulting from the electron-phonon interaction
has at low temperatures logarithmic singularities (in the real part) and discontinuities (in the imaginary one) at the frequencies corresponding to the opening of the inelastic channels.
These singularities  generate discontinuities and
logarithmic divergences (as a function of the bias voltage)  in the  low-temperature differential conductance
around the inelastic thresholds.  The self energy also depends on the population of the vibrational modes. The case of a vibrating  free junction   (not coupled to a thermal bath), where the phonon population is determined by the bias voltage  is examined. We compare the resulting zero-temperature differential conductance with the one obtained for equilibrated phonons, and find that the difference is larger the larger is the bare transmission of the junction and the product  of the electron dwell time on the junction
with  the phonon frequency.

\end{abstract}



\maketitle

Electrons passing through a small molecule  can change its
quantum state and this  usually requires a finite energy transfer from the
transport electron, yielding interesting structures in the I-V characteristics.
This rich pattern
depends on featuress such as  the
equilibration time of the vibrations compared to the typical time
between consecutive electrons passing through the junction,   or
whether the electrons can pump more and more excitations into the
vibrational states.
Single-molecule junctions based on direct bonding of a small
molecule between two metallic electrodes  seem  by now rather
established experimentally.
\cite{REED,reichert,zhitenev,kubatkin,QIU,kushmerick,pasupathy,dekker,djukic,FC1,FC2}
There are also quite a number of theoretical studies, focusing on various regimes of the relevant parameters. \cite{LUNDIN,MAC,flens,AJI,THOSS,MITRA,FELIX,CHEM,ryndyk,vega,AKIKO,paulson,NITZAN,gogolin,SCHMIDT,munich}  In particular, the modification of the differential conductance at the opening of the inelastic channel has been an issue of considerable interest (an extended discussion  may be found in Ref. ~\onlinecite{we}).

An electron crossing  the molecular bridge can do so with or without  changing the vibrational
excitation state of the molecule. At low temperatures, the first
inelastic channel comes in when the bias voltage,   $V$,      exceeds $\hbar
\omega^{}_{0}/e$, where $\omega_{0}$ is the normal  frequency of
the junction. This however does not necessarily imply an increase
of the total conductance,  since the elastic conduction channel is
modified as well. Technically, the low-temperature  conductance associated with the
opening of the inelastic channel  stems from two sources. The first is the imaginary part of
the junction self energy, induced by   the interaction with  the
oscillator. This function develops discontinuities, so that there appear additional contributions to the conductance   only for $eV>\hbar\omega_{0}$.  
Discontinuities in the imaginary part of the self energy imply
logarithmic divergences in its real part, via
 the Kramers-Kronig relations. \cite{gogolin,we}
As a result, the low-temperature differential conductance develops
logarithmic singularities (as a function of the bias voltage)
around the inelastic thresholds. \cite{MITRA,flens,gogolin} The second contribution is due to a nonequilibrium population of the vibrations.  At zero temperature, phonons can be excited only when the energy of the electrons (i.e., $eV$) exceeds $\omega_{0}$, leading to an additional modification of the conductance at these voltages.

In a recent publication \cite{we} we have presented a detailed
calculation of the differential conductance and analyzed its behavior at low temperatures. Our
calculation implicitly assumed that the molecular junction is in a
good contact with a thermal bath, such that the vibrations follow
the Bose-Einstein distribution, with the same temperature as the
two leads.  In particular we
have considered the conditions for the conductance to increase or
decrease at the channel opening.
In this short
communication we repeat that analysis for the case of a free
junction (i.e.,  the oscillator is not coupled to a heat bath), for which at low temperatures the phonon population is
determined by the bias voltage. This case is analogous to the magnetization of a Kondo ion out of equilibrium. \cite{ACHIM}

In this note we use the same Hamiltonian and the same notations as in our previous paper. \cite{we}
To lowest order in the electron-phonon interaction on the dot, $\gamma$, the scattering-down
rate (de-exciting the vibration) can be derived from the golden-rule, 
\begin{align}
w^{}_{\rm down}&=\gamma^{2}_{}\frac{\pi}{2}\int d\omega {\cal N}(\omega ){\cal N}(\omega +\omega^{}_{0})\nonumber\\
&\times\sum_{\alpha, \alpha ' =L,R}f^{}_{\alpha}(\omega )\Bigl (1-f^{}_{\alpha '}(\omega +\omega^{}_{0})\Bigr )\ ,\label{OUT}
\end{align}
and the scattering-up rate, $w_{\rm up}$, is obtained from Eq.
(\ref{OUT}) by swapping $\omega_{0}$ and $-\omega_{0}$. Here,
$f_{\alpha}(\omega )=(\exp[\beta (\omega -\mu_{\alpha})]+1)^{-1}$
is the Fermi distribution in the left ($\alpha =L$) or the right
($\alpha =R$) lead, in which the chemical potential is
$\mu_{\alpha}$, and ${\cal N}(\omega )$ is the bare density of
states on the dot,
\begin{align}
{\cal N}(\omega )=\frac{\Gamma^{}_{0}/\pi}{\omega^{2}+\Gamma^{2}_{0}}\ .\label{DOS}
\end{align}
It is assumed  that the dot, modeled by a single level of energy $\epsilon_{}$, is coupled symmetrically to the two leads,  with $\Gamma_{0}$ being the bare resonance width and $eV=\mu_{L}-\mu_{R}$, with $\mu_{L}>\mu_{R}$.
All electronic frequencies (using $\hbar =1$) are measured from $\epsilon -\mu$, where $\mu=(\mu_{L}+\mu_{R})/2$. 
The
kinetic equation for the vibration population, $N_{\rm ho}$, is then
 (see also Ref. ~\onlinecite{MITRA})
\begin{align}
&\frac{dN^{}_{\rm ho}(t)}{dt}=-N^{}_{\rm ho}(t)w^{}_{\rm
down}+[1+N^{}_{\rm ho}(t)]w^{}_{\rm up}\ .\label{KE}
\end{align}
 The scattering-up rate is due to all processes by which an electron can excite an
 oscillator mode, losing the energy $\omega_{0}$ in the process, and moving over
 from the left (right) lead back into the left (right) lead, or from the left
 (right) lead into the right (left) one. Hence, at zero temperature, $w_{\rm up}\neq 0$ 
 only when the bias voltage exceeds
 the frequency $\omega_{0}$, resulting from 
 the $L\rightarrow R$ process.
On the other hand, $w_{\rm down}$
results
from all four
processes in which the electron gains the energy $\omega_{0}$.
However, at zero temperature
and when $eV\geq\omega_{0}$
the
$R\rightarrow L$ process {\em ceases} to contribute. It follows from Eq. (\ref{KE})  that the   stationary vibration  population,  
\begin{align}
N^{}_{\rm ho}=\frac{w^{}_{\rm up}}{w^{}_{\rm down}-w^{}_{\rm up}}\
,
\end{align}
{\em is independent of the electron-phonon coupling}, \cite{ACHIM} and vanishes at zero temperature as long as $eV\leq\omega_{0}$.

\begin{figure}[ h]
\includegraphics[width=7cm]{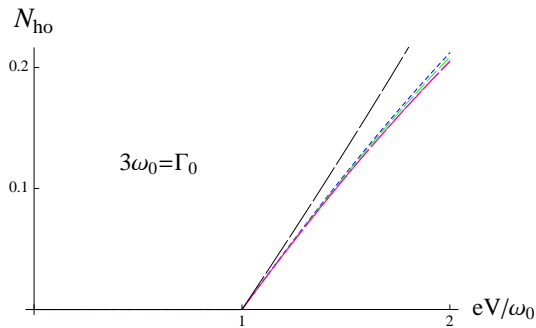}
\caption{The population of the oscillator modes, $N_{\rm ho}$,  at zero temperature, as a function of $eV/\omega_{0}$,  for various values of the bare transmission: 
${\cal T}$=0.2 (smallest dash size), 4/17, 0.3, 4/9, and 0.9 (largest dash  size). Here $\omega_{0}/\Gamma_{0}=1/3$.
}\label{FIGNHOa}
\end{figure}

\begin{figure}[h ]
\includegraphics[width=7cm]{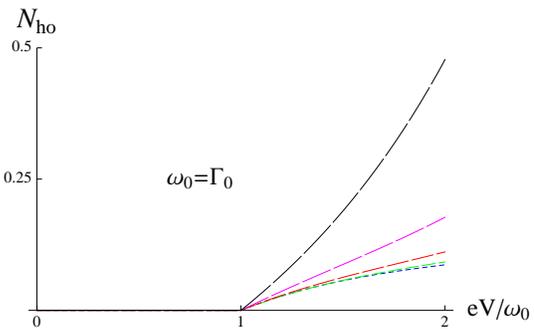}
\caption{The same as in Fig. \ref{FIGNHOa}, for $\omega_{0}=\Gamma_{0}$.
}\label{FIGNHOb}
\end{figure}

\begin{figure}[ h]
\includegraphics[width=7cm]{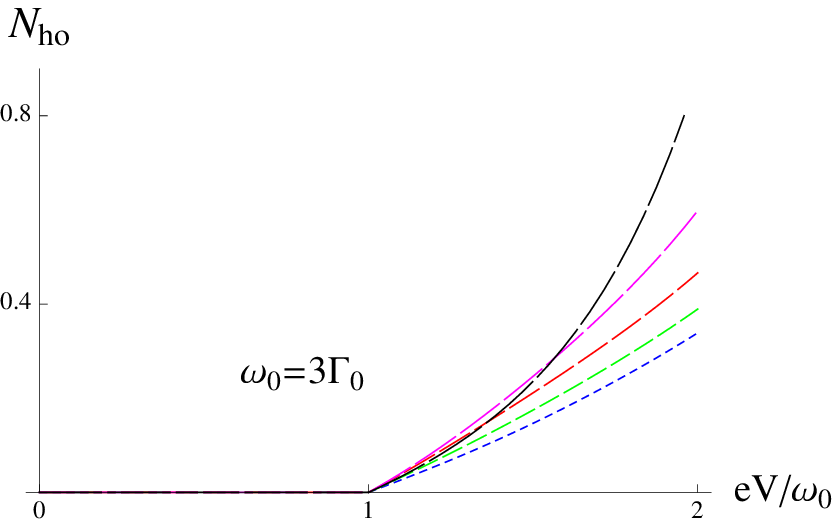}
\caption{The same as in Fig. \ref{FIGNHOa}, for $\omega_{0}/\Gamma_{0}=3$.
}\label{FIGNHOc}
\end{figure}

Figures \ref{FIGNHOa}, \ref{FIGNHOb}, and \ref{FIGNHOc} depict the nonequilibrium phonon population
at zero temperature (see also Ref. ~\onlinecite{MITRA}). The
curves 
are for various values of the bare
transmission, ${\cal
T}=\Gamma_{0}^{2}/(\mu^{2}+\Gamma^{2}_{0})$,   of the junction, 
The
population usually increases with ${\cal T}$ and with $V$, and its
 magnitude increases with the ratio
$\omega_{0}/\Gamma_{0}$. \cite{note} This can be understood qualitatively: 
since $eV-\omega_{0}$ is the driving force pushing the population out of equilibrium, $N_{\rm ho}$ increases with the voltage. The increase with ${\cal T}$ is due to the fact that the dwell time  on the dot, $\tau_{d}$, increases with ${\cal T}$ from being very short off resonance to becoming $\sim 1/\Gamma_{0}$ around the resonance. As emphasized in Ref. ~\onlinecite{we}, to effectively excite the phonon, $\tau_{d}$ should be longer than the response
time of the oscillator (about $\omega_{0}^{-1}$), i.e.,
$\Gamma_{0}< \omega_{0}$.
Another interesting issue
is that the  time interval between successive electrons passing
the junction,  $\tau_{c}\sim e/I\sim
1/(eV{\cal T})$, decreases as ${\cal T}$ or $eV$ are enhanced.
This too will cause
$eV$ to affect more significantly the population
at higher values of the bare transmission.
However, when  $\tau_{c} < \tau_{d}$, Pauli constraints between consecutive electrons on the dot should come into play.

When $\Gamma_{0} \gg \omega_{0}$, one may estimate the vibration population at finite temperatures by assuming  that  the bare density of states, [see Eq. (\ref{DOS})],  is a constant. Then the stationary population becomes
\begin{align}
1+2N^{}_{\rm ho}&=\frac{1}{2}{\rm coth}\frac{\beta\omega^{}_{0}}{2}
+\frac{1}{4}\Bigl (1+\frac{eV}{\omega^{}_{0}}\Bigr ){\rm coth}\frac{\beta(\omega^{}_{0}+eV)}{2}\nonumber\\
&
+\frac{1}{4}\Bigl (1-\frac{eV}{\omega^{}_{0}}\Bigr ){\rm coth}\frac{\beta (\omega_{0}^{}-eV)}{2}\ ,
\end{align}
independent of the bare resonance width $\Gamma_{0}$ and of the
band width of the leads. This resembles the magnetization of a Kondo ion out of equilibrium, \cite{ACHIM}
with the oscillator frequency $\omega_{0}$ playing the role of the
applied magnetic field.
 (That magnetization is the inverse of
$1+2N_{\rm ho}$.)

The current flowing through the junction can be presented in the form [see, e.g., Ref. ~\onlinecite{we}]
\begin{align}
I=e\Gamma^{}_{0}\int\frac{d\omega}{2\pi}{\rm Im}G^{a}_{00}(\omega )\Bigl (f^{}_{L}(\omega )-f^{}_{R}(\omega )\Bigr )\ ,\label{CURRENT}
\end{align}
where ${\rm Im}G^{a}_{00}/\pi$ is the density of states on the dot, fully dressed by the interaction with the vibrations. The (advanced) Green function on the junction is
\begin{align}
G^{a}_{00}(\omega )=\frac{1}{\omega -i\Gamma^{}_{0}-\Delta\epsilon^{}_{0}-\Sigma^{a}_{\rm ho}(\omega )}\ ,
\end{align}
where $\Delta\epsilon_{0}$ is the polaron shift,
\begin{align}
\Delta\epsilon^{}_{0}=-\frac{\gamma^{2}}{\omega^{}_{0}}\int d\omega {\cal N}(\omega )\Bigl (f^{}_{L}(\omega )+f^{}_{R}(\omega )\Bigr )\ ,\label{POL}
\end{align}
and $\Sigma_{\rm ho}$ is the self-energy due to the electron-phonon processes. Up to second-order in the electron-phonon coupling it reads \cite{we}
\begin{align}
\Sigma^{a}_{\rm ho}(\omega )&=\frac{\gamma^{2}}{2}
\int   d\omega '{\cal N}(\omega ')
\Biggl  (\frac{2N^{}_{\rm ho}+\sum_{\alpha=L,R }(1-f^{}_{\alpha}(\omega '))} {\omega -\omega^{}_{0}-\omega '- i 0^{+}}\nonumber\\
&+
\frac{2N^{}_{\rm ho}+\sum_{\alpha=L,R }f^{}_{\alpha}(\omega ')}{\omega +\omega^{}_{0}-\omega '- i0^{+}}\Biggr )\ .\label{SIGHO}
\end{align}
Employing Eqs. (\ref{POL}) and  (\ref{SIGHO}) in Eq.
(\ref{CURRENT}), we have derived the differential conductance of the bridge,  ${\rm G}$, at zero temperature, and to second order in the coupling with the phonons.

\begin{figure}[ h]
\includegraphics[width=7cm]{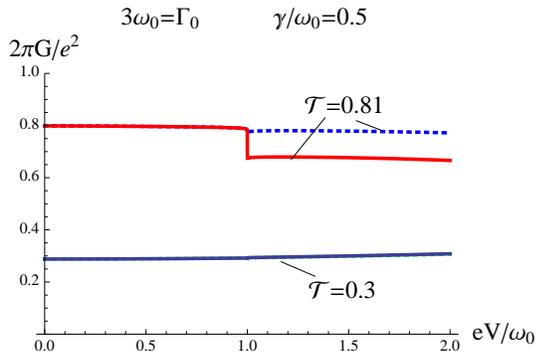}
\caption{The  zero-temperature differential conductance, as a function of $eV/\omega_{0}$, for two values of the bare transmission, marked on the figure. The full lines show the conductance with  the nonequilibrium phonon population, the dashed curves are with the equilibrium one. Here $\omega_{0}/\Gamma_{0}=1/3$.}\label{Ga}
\end{figure}

\begin{figure}[ h]
\includegraphics[width=7cm]{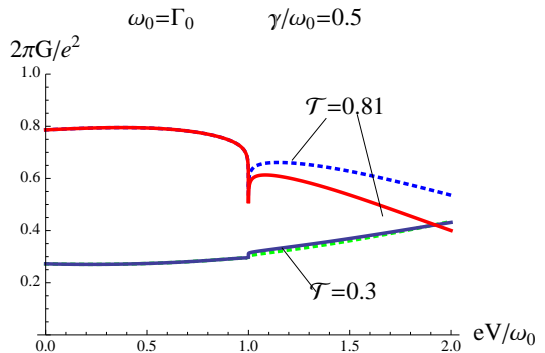}
\caption{The same as in Fig. \ref{Ga}, for $\omega_{0}=\Gamma_{0}$.}\label{Gb}
\end{figure}

\begin{figure}[ hbtp]
\includegraphics[width=7cm]{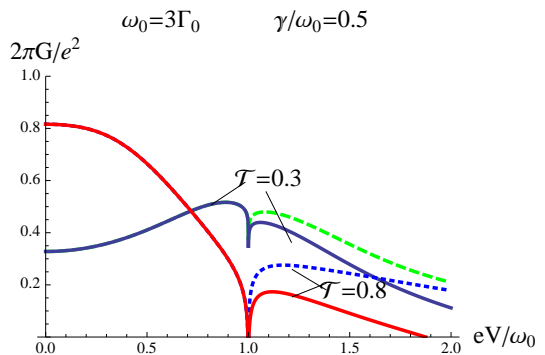}
\caption{The same as in Fig. \ref{Ga}, for $\omega_{0}=3\Gamma_{0}$. }\label{Gc}
\end{figure}

Figures \ref{Ga},  \ref{Gb}, and \ref{Gc} show several examples of the dependence of  the differential conductance on the bias voltage and on the other parameters of the junction. The first figure corresponds to the case where the junction is tightly bound to the leads, and hence the dwell time of the electrons is rather short. Then, at relatively low values of the bare transmission (${\cal T}=0.3$ in our example) there is no discernible  modification in the conductance which is about the same either for nonequilibrium phonon  population or for the equilibrium one.  However, for higher values of the bare transmission (${\cal T}=0.81$) the step-down feature of the conductance at threshold for inelastic tunneling is enhanced for nonequilibrium phonons as compared to the equilibrated ones. 

From the few examples portrayed in Figs. \ref{Ga}-\ref{Gc} we see that when the bare transmission of the junction is high, then the conductance in the presence of nonequilibrium phonons is lower than the one pertaining to the case of equilibrated vibrations. For low bare transmissions the difference is rather small. One  notes that  the
logarithmic singularity associated with the {\em
real} part of the self energy at  the channel opening, which has
been previously discussed \cite{flens,gogolin,we} for an
equilibrium population of the phonons, is still manifested also
when these excitations are out of equilibrium, for high enough values of
the bare transmission of the junction. 

The fact that the changes in the differential conductance which are associated with the type of the vibration population, $N_{\rm ho}$,  are more pronounced  for the larger  $\omega_{0}/\Gamma_{0}$   ratio
is connected with the actual value of $N_{\rm ho}$, see Figs. \ref{FIGNHOa}-\ref{FIGNHOc}.  The electrons pump more and more excitations into the higher vibrational states as the bias voltage increases and this pumping is more effective as the dwell time exceeds considerably the response time of the oscillator.

\begin{acknowledgments}
We thank A. Rosch and P. W\"{o}lfle for  crucial discussions. This work was supported by the German Federal Ministry of
Education and Research (BMBF) within the framework of the
German-Israeli project cooperation (DIP), by the US-Israel
Binational Science Foundation (BSF), by the Israel Science
Foundation (ISF) and by its Converging Technologies Program.
\end{acknowledgments}


\begin{thebibliography}{999}




\bibitem{REED}

M. A. Reed, C. Zhou, C. J. Muller, T. P. Burgin, and J. M. Tour, Science {\bf 278}, 252 (1997); J. Chen, M. A. Reed, A. M. Rawlett, and J. M. Tour, {\it ibid} {\bf 286}, 1550 (1999).


\bibitem{reichert}


J. Reichert, R. Ochs, D. Beckmann, H. B. Weber, M. Mayor, and H. v. Lohneysen, Phys. Rev. Lett. {\bf 88}, 176804 (2002).

\bibitem{zhitenev}

N. B. Zhitenev, H. Meng, and Z. Bao, Phys. Rev. Lett. {\bf 88}, 226801 (2002).

\bibitem{kubatkin}

S. Kubatkin, A. Danilov, M. Hjort, J. Cornil, J. Bredas, N. Stuhr-Hansen, P. Hedegard, and T. Bjornholm, Nature (London) {\bf 425}, 698 (2003).



\bibitem{kushmerick}


J. G. Kushmerick, J. Lazorcik, C. H. Patterson, R. Shashidhar, D. S. Seferos, and G. C. Bazan, Nano Lett. {\bf 4}, 639 (2004).


\bibitem{QIU}

X. H. Qiu, G. V. Nazin, and W. Ho, Phys. Rev. Lett. {\bf 92}, 206102 (2004).

\bibitem{pasupathy}

H. Park, J. Park, A. K. L. Lim, E. H. Anderson, A. P. Alivisatos, and P. L. McEuen, Nature (London) {\bf 407}, 57 (2000);
A. N. Pasupathy, J. Park, C. Chang, A. V. Soldatov, S. Lebedkin, R. C. Bialczak, J. E. Grose, L. A. K. Donev, J. P. Sethna, D. C. Ralph, and P. L. McEuen, Nano Lett. {\bf 5}, 203 (2005).

\bibitem{dekker}

B. J. LeRoy, S. G. Lemay, J. Kong, and C. Dekker, Nature (London) {\bf 432}, 371 (2004).

\bibitem{djukic}

R. H. M. Smit, Y. Noat, C. Untiedt, N. D. Lang, M. C. van Hemert, and J. M. van Ruitenbeek, Nature (London) {\bf 419}, 906 (2002);
D. Djukic, K. S. Thygesen, C. Untiedt, R. H. M. Smit, K. W. Jacobsen, and J. M. van Ruitenbeek, Phys. Rev. B. {\bf 71}, 161402(R) (2005);
W. H. A. Thijssen, D. Djukic, A. F. Otte, R. H. Bremmer, and J. M. van Ruitenbeek, Phys. Rev. Lett. {\bf 97}, 226806 (2006);
O. Tal, M. Krieger, B. Leerink, and J. M. van Ruitenbeek,
Phys. Rev. Lett. {\bf 100}, 196804 (2008);
M. Kiguchi, O. Tal, S. Wohlthat, F. Pauly, M. Krieger, D. Djukic, J. C. Cuevas, and
J. M. van Ruitenbeek, Phys. Rev. Lett. {\bf 101}, 046801 (2008).


\bibitem{FC1}

S. Sapmaz, P. Jarillo-Herrero, Ya. M. Blanter, C.  Dekker, and H. S. J. van der Zant, Phys. Rev. Lett. {\bf 96}, 026801 (2006).





\bibitem{FC2}

R. Leturcq, C. Stampfer, K. Interbitzin, L. Durrer, C. Heirold, E. Mariani, M. G. Schultz, F. von Oppen, and K. Ensslin, Nature Physics {\bf 5}, 327 (2009).




\bibitem{LUNDIN}

U. Lundin and R. H. McKenzie, Phys. Rev. B {\bf 66}, 075303 (2002).



\bibitem{MAC}
K. D. McCarthy, N. Prokof'ev, and M. T. Tuominen, Phys. Rev. B {\bf 67}, 245415 (2003).



\bibitem{flens}
K. Flensberg, Phys. Rev. B {\bf 68}, 205323 (2003).




\bibitem{AJI}
V. Aji, J. E. Moore, and C. M. Varma,  Int. J. Nano. {\bf 3}, 255 (2004).



\bibitem{THOSS}

M. Cizek, M. Thoss, and W. Domcke, Phys. Rev. B {\bf 70}, 125406 (2004).



\bibitem{MITRA}

A. Mitra, I. Aleiner, and A. J. Millis, Phys. Rev. B {\bf 69}, 245302  (2004); Phys. Rev. Lett. {\bf 94}, 076404 (2005).






\bibitem{FELIX}

J. Koch and F. von Oppen, Phys. Rev. Lett. {\bf 94}, 206804 (2005).


\bibitem{CHEM}

J. K. Viljas, J. C. Cuevas, F. Pauly, and M. H\"{a}fner, Phys. Rev. B {\bf 72}, 245415 (2005).

\bibitem {ryndyk}

D. A. Ryndyk, M. Hartung, and G. Cuniberti, Phys. Rev. B {\bf 73}, 045420 (2006).

\bibitem{vega}

L. de la Vega, A. Martin-Rodero, N. Agrait, and A. Levy Yeyati, Phys. Rev. B {\bf 73}, 075428 (2006).



\bibitem{AKIKO}


A. Ueda and M. Eto, Phys. Rev. B {\bf 73}, 235353 (2006).







\bibitem{paulson}
M. Paulsson, T. Frederiksen, and M. Brandbyge, Phys. Rev. B {\bf 72}, 201101(R) (2005); T. Frederiksen, N. Lorente, M. Paulsson, and M. Brandbyge, {\it ibid.} {\bf 75}, 235441 (2007).









\bibitem{NITZAN}

M. Galperin, M. A. Ratner, and A. Nitzan, J. Chem. Phys. {\bf 121}, 11965 (2004); J. Phys.: Condens. Matter {\bf 19}, 103201 (2007).



\bibitem{gogolin}
R. Egger and A. O. Gogolin, Phys. Rev. B {\bf 77}, 113405 (2008).




\bibitem{SCHMIDT}

T. L. Schmidt, and A. Komnik, Phys. Rev. B {\bf 80}, 041307(R) (2009); R. Avriller and A. Levy Yeyati, Phys. Rev. B {\bf 80}, 041309(R) (2009);
F. Haupt, T. Novotn\`{y}, and W. Belzig, Phys. Rev. Lett. {\bf 103}, 136601 (2009).








\bibitem{munich}

R. H\"{a}rtle, C. Benesch, and M. Thoss, Phys. Rev. Lett. {\bf 102},
146801 (2009);


\bibitem{we}



O. Entin-Wohlman, Y. Imry, and A. Aharony,
Phys. Rev. B {\bf  80}, 035417 (2009), and references therein.


\bibitem{ACHIM}
A. Rosch, private communication; see also
A. Rosch, J. Paaske, J. Kroha, and P. W\"{o}lfle, Phys. Rev. Lett. {\bf 90},
076804 (2003); O. Parcollet and C. Hooley, Phys. Rev. B {\bf 66}, 085315 (2002).





\bibitem{note} In some cases, $N_{\rm ho}$ does not increase
monotonically with $V$ or with ${\cal T}$. However, this is a
small effect, occuring mainly at relatively small $V$.






\end{thebibliography}
\end{document}